\documentclass[12pt]{article}
\usepackage{geometry}                
\usepackage{graphicx}
\usepackage{amssymb}
\usepackage{float}
\usepackage{url,hyperref}

\title{Bell's theorem is an exercise in the statistical theory of causality}
\author{Richard D. Gill\\  \it Mathematical Institute, Leiden University}
\date{\normalsize  Version 2: 20 March, 2023
\\ \small Several typos corrected. ~~\texttt{arXiv.org:2211.05569} }                                           

\begin{document}
\maketitle

\begin{quote}
{\bf Abstract.} In this short note, I derive the Bell-CHSH inequalities as an elementary result in the present-day theory of statistical causality based on graphical models or Bayes' nets, defined in terms of DAGs (Directed Acyclic Graphs) representing direct statistical causal influences between a number of observed and unobserved random variables. I show how spatio-temporal constraints in loophole-free Bell experiments, and natural classical statistical causality considerations, lead to Bell's notion of local hidden variables, and thence to the CHSH inequalities. The word ``local'' applies to the way that the chosen settings influence the observed outcomes. The case of contextual setting-dependent hidden variables (thought of as being located in the measurement devices and dependent on the measurement settings) is automatically covered, despite recent claims that Bell's conclusions can be circumvented in this way.
\end{quote}

In this short note I will derive the Bell-CHSH inequalities as an exercise in the modern theory of causality based on Bayes' nets: causal graphs described by DAGs (directed acyclic graphs). The note is written in response to a series of papers by M. Kupczynski \cite{MK2, MK3, MK4, MK1} in which that author claims that Bell-CHSH inequalities cannot be derived (the author in fact writes \emph{may} not be derived) when one allows contextual setting-dependent hidden variables thought of as being located in the measurement devices and with probability distributions dependent on the local setting. The result has of course been
known for a long time, but it seems worth writing out in full for the benefit of ``the probabilistic opposition'' as a vociferous group of critics of Bell's theorem like to call themselves.

\begin{figure}[H]
\noindent\centerline{\includegraphics[width=0.95\textwidth]{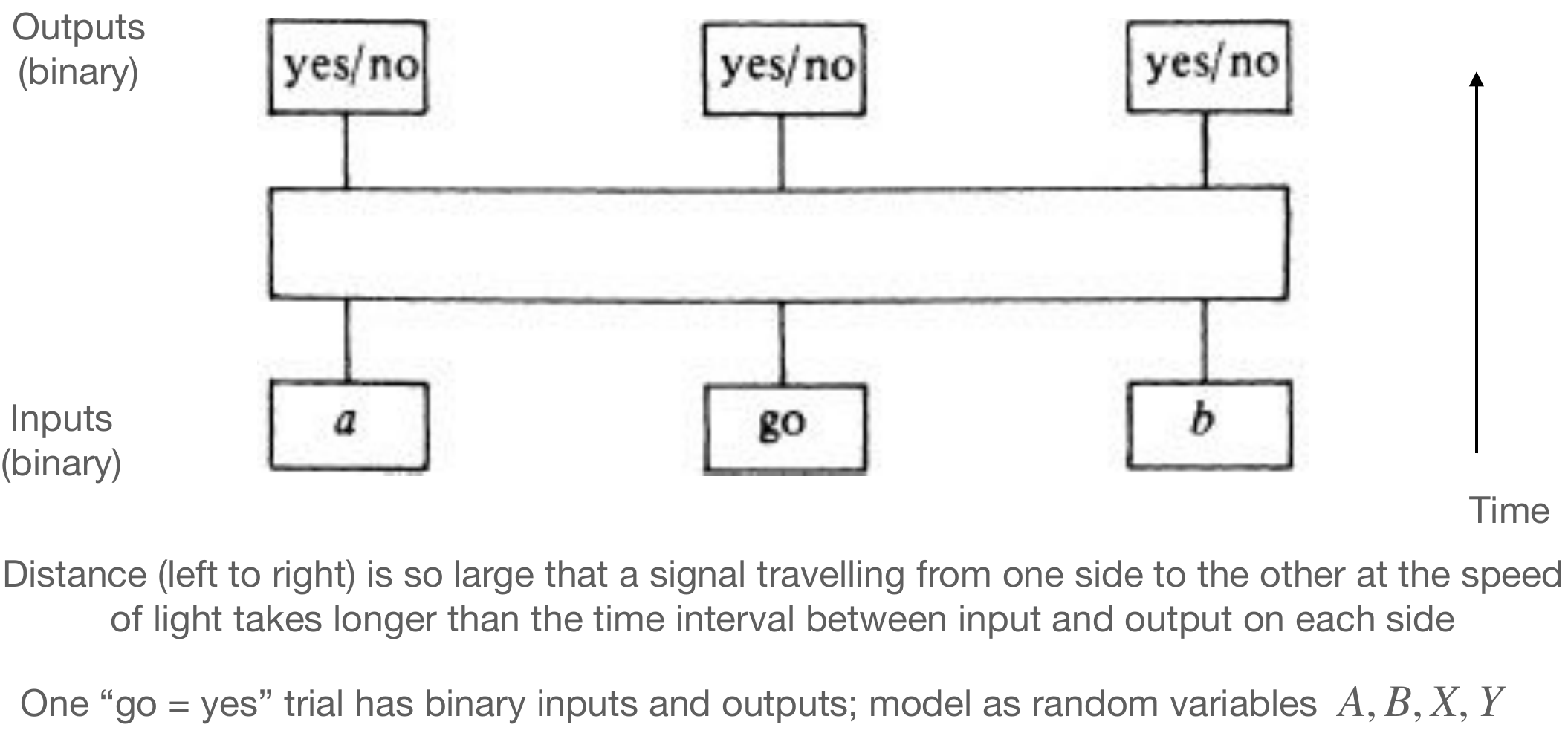}}
\caption{Spatio-temporal disposition of one trial of a Bell experiment.\\ \small (Figure 7 from J.S. Bell (1981), “Bertlmann’s socks and the nature of reality”)}
\label{fig1}
\end{figure}

\begin{figure}[H]
\noindent\centerline{\includegraphics[width=0.95\textwidth]{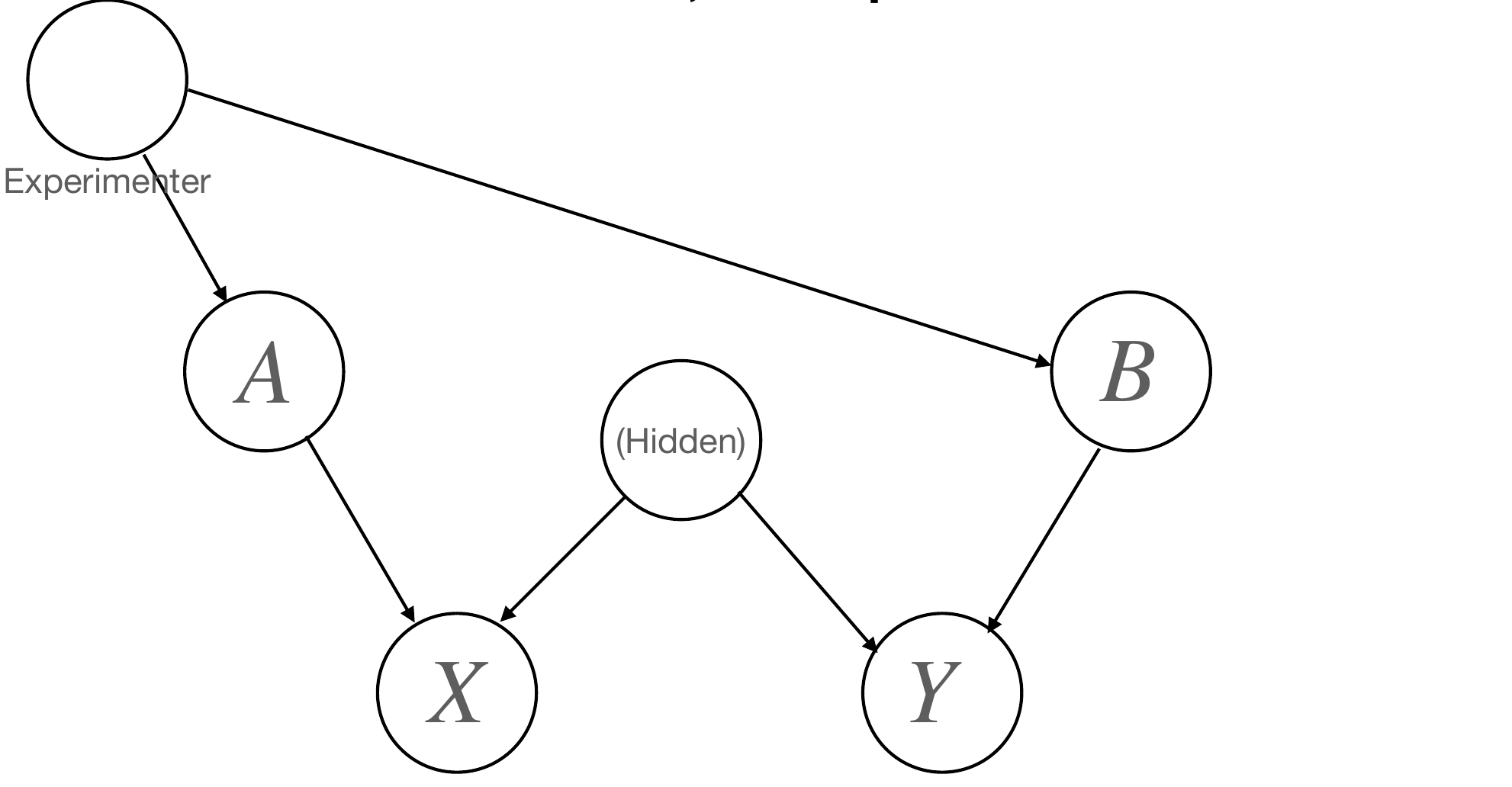}}
\caption{Graphical model of one trial of a Bell experiment}
\label{fig2}
\end{figure}

Figure \ref{fig1} gives us the physical background and motivation for the causal model described in the DAG of Figure \ref{fig2}. How that is arranged (and it can be arranged in different ways) depends on Alice and Bob's assistant, Charlie, at the intermediate location in Figure \ref{fig1}. There is no need to discuss his or her role in this short note. Very different arrangements can lead to quite different kinds of experiments, from the point of view of their realisation in terms of quantum mechanics.
Figure \ref{fig1} is meant to describe the spatio-temporal layout of one \emph{trial} in a long run of such trials of a fairly standard loophole-free Bell experiment. At two distant locations, Alice and Bob each insert a setting into an apparatus, and a short moment later, get to observe an outcome. Settings and outcomes are all binary. One may imagine two large machines each with a switch on it that can be set to position ``up'' or ``down''; one may imagine that it starts in some neutral position. A short moment later, a light starts flashing: it could be red or green. Alice and Bob each write down their setting and their outcome. This is repeated many times. The whole thing is synchronised (with the help of Charlie at the central location). The trials are numbered, say from $1$ to $N$, and occupy short \emph{time-slots} of fixed length. The arrangement is such that Alice's outcome has been written down before a signal carrying Bob's setting could possibly reach Alice's apparatus, and vice versa.

As explained, each trial has two binary inputs or settings, and two binary outputs or outcomes. I will denote them using the language of classical probability theory by random variables $A, B, X, Y$ where $A,B\in\{1, 2\}$ and $X, Y\in\{-1, +1\}$. A complete experiment corresponds to a stack of $N$ copies of this graphical model, ordered by time. We will not make any assumptions whatsoever (for the time being) about independence or identical distributions. The experiment does generate an $N \times 4$ spreadsheet of 4-tuples $(a, b, x, y)$.  The settings $A$, $B$ should be thought of merely as labels (categorical variables); the outcomes $X,Y$ will be thought of as numerical. In fact, we will derive inequalities for the four \emph{correlations} $\mathbb E(XY | A=a, B= b)$ for one trial.
In Figure \ref{fig2}, the nodes labelled $A$, $B$, $X$, $Y$ correspond to the four observed binary variables. The other two nodes annotated ``Experimenter'' and ``(Hidden)'' correspond to factors leading to the statistical dependence structure of $(A, B, X, Y)$ of two kinds. On the one hand, the experimenter externally has control of the choice of the settings. In some experiments they are intended to be the results of external, fair coin tosses. Thus the experimenter might try to achieve that $A$ and $B$ are statistically independent and completely random. The important thing is the aim to have the mechanism leading to selection of two settings statistically independent of the physics of what is going on inside the long horizontal box of Figure \ref{fig1}. That mechanism is unknown and unspecified. In the physics literature, one uses the phrase ``hidden variables'', and they are thought to be those aspects of the initial state of all the stuff inside the long box which leads in a quasi-deterministic fashion to the actually observed measurement outcomes. The model therefore represents a classical physical model, classical in the sense of pre-quantum theory, and one in which experimental settings can be chosen in a statistically independent manner from the parameters of the physical processes, essentially deterministic, which lead to the actually observed measurement outcomes at the two ends of the long box.

Thus we are making the following assumptions. There are two statistically independent random variables (not necessarily real-valued -- they may take values in any measure spaces whatsoever), which I will denote by $\Lambda_E$ and $\Lambda_H$, such that the probability distribution of $(A, B, X, Y)$ can be simulated as follows. First of all, draw outcomes $\lambda_E$ and $\lambda_H$, independently, from any two probability distributions over any measure spaces whatsoever. Next, given $\lambda_E$, draw outcomes $a, b$ from any two probability distributions on $\{1, 2\}$, depending on $\lambda_E$. Next, given $a$ and $\lambda_H$, draw $x\in \{-1, +1\}$ from some probability distribution depending on those two parameters, and similarly, independently, draw $y \in\{-1, +1\}$ from some probability distribution depending on $b$ and $\lambda_H$\footnote{In this Kolmogorovian mathematical framework there is a ``hidden'' technical assumption of measurability. It can be avoided, see the author's 2014 paper ``Statistics, Causality and Bell's Theorem'', published in the journal \emph{Statistical Science} and also available on \texttt{arXiv.org}. The assumption of $N$ \emph{independent and identically distributed} copies of this picture can be avoided too.}.

Thus, $\Lambda_H$ is the hidden variable responsible for possible statistical dependence between $X$ and $Y$ given $A$ and $B$. 

In the theory of graphical models, one knows that such models can be thought of as deterministic models, where the random variable connected to any node in the DAG is a deterministic function of the variables associated with nodes with direct directed links to that node, together with some independent random variable associated with that node. In particular therefore, in obvious notation,
$$X = f(A, \Lambda_H, \Lambda_X),$$
$$Y = g(B, \Lambda_H, \Lambda_Y),$$
where $\Lambda := (\Lambda_H, \Lambda_X, \lambda_Y)$ is statistically independent of $(A, B)$, the three components of $\Lambda$ are mutually independent of one another, and $f$, $g$ are some functions. We can now redefine the functions $f$ and $g$ and rewrite the last two displayed equations as
$$X = f(A, \Lambda),$$
$$Y = g(B, \Lambda),$$
where $f$, $g$ are some functions and $(A, B)$ is statistically independent of $\Lambda$. This is what Bell called a local hidden variables model. It is absolutely clear that Kupczynski's notion of a probabilistic contextual local causal model is of this form. It is a special case of the \emph{non-local} contextual model
$$X = f(A, B, \Lambda),$$
$$Y = g(A, B, \Lambda),$$
in which Alice's outcome can also depend directly on Bob's setting, and vice versa.

Kupczynski claims that Bell inequalities cannot (or may not?) be derived for his model. But that is easy. Thanks to the assumption that $(A, B)$ is statistically independent of $\Lambda$, one can define four random variables $(X_1, X_2, Y_1, Y_2)$ as
$$X_a = f(a, \Lambda)$$
$$Y_b= g(b, \Lambda).$$
These four have a joint probability distribution by construction, and take values in $\{-1, +1\}$. By the usual simple proof, all Bell-CHSH inequalities hold for the four correlations $\mathbb E(X_a Y_b)$. But each of these four correlations is identically equal to the ``experimentally accessible'' correlation $\mathbb E(XY \mid A=a, B = b)$; for all $a, b$
$$\mathbb E(X_a Y_b) = \mathbb E(XY \mid A=a, B = b)$$
while $$-2 ~ \le ~ \mathbb E(X_1Y_1) - \mathbb E(X_1Y_2) - \mathbb E(X_2Y_1) - \mathbb E(X_2Y_2) ~ \le ~ +2$$
and similarly for the comparison of each of the other three correlations with the sum of the others.

The whole argument also applies (with a little more work) to the case when the outcomes lie in the set $\{-1, 0, +1\}$, or even in the interval $[-1, +1]$. An easy way to see this is to interpret values in $[-1, +1]$ taken by $X$ and $Y$ not as the actual measurement outcomes, but as their expectation values given relevant settings and hidden variables. One simply needs to add to the already hypothesized hidden variables further independent uniform $[0, 1]$ random variables to realise a random variable with given conditional expectation in $[-1, 1]$ as a function of the auxiliary uniform variable. The function depends on the values of the conditioning variables. Everything stays exactly as local and contextual as it already was.

\raggedright

\end{document}